# THE EMBEDDING OF GENERAL RELATIVITY IN FIVE-DIMENSIONAL CANONICAL SPACE: A SHORT HISTORY AND A REVIEW OF RECENT PHYSICAL PROGRESS

Paul S. Wesson

Department of Physics and Astronomy, University of Waterloo, Waterloo, Ontario N2L 3G1, Canada

Abstract: Einstein theory can be embedded in Kaluza-Klein theory, and in particular all 4D vacuum solutions can be embedded in 5D (pure) canonical space where spacetime is independent of the extra coordinate. The uniqueness of 5D canonical space is quickly reproven, and its history and implications briefly reviewed. For 4D anti-deSitter space, the 5D space involves a wave running around spacetime, with an associated particle that has a local value of the cosmological 'constant' proportional to the square of the mass.

Correspondence: mail to University of Waterloo, email = psw.papers@yahoo.ca

1. Introduction

It is a Theorem, that all solutions of 4D general relativity in vacuum but with a cosmological constant can be locally embedded in 5D (pure) canonical space. This applies to the Schwarzschild solution, more complicated solutions like those for a spinning black hole and gravitational waves, and 4D deSitter space. The last is particularly simple, and will be used below to illustrate the applications of 5D embeddings to physics. In fact, 5D embeddings inform about the geometrical 'environment' of 4D solutions, helping to understand what happens in spacetime and pointing towards new results in the extra dimension. The subject of embeddings is presently very active, so it may be opportune to give a brief history and to review recent discoveries.

This is done in Sections 2 and 3 following. The present account is in the nature of a discussion paper or status report. The adage says that those ignorant of history are sometimes doomed to repeat it. However, those conversant with embeddings may wish to proceed to the new physical applications of Section 3. The last section is a conclusion.

2. A Brief History of the 4D / 5D Canonical Embedding

Most people know that Einstein's general theory of relativity was effectively extended from four to five dimensions by Kaluza in 1920, as a means of unifying gravitation and electromagnetism [1]. He did this essentially by identifying the off-diagonal elements of a 5x5 metric tensor with the Maxwell potentials of electromagnetism. This approach was modified by Klein in 1926, as a means of explaining the quantization of electric charge [2]. Both men suppressed the diagonal fifth component of the metric tensor, which represents a scalar field now believed by some workers to be related to the Higgs field of quantum field theory and connected to particle mass [3]. Leaving this aside, it was noted already by Kasner in 1921 that most curved 4D solutions of general relativity could not be embedded in a *flat* 5D space [4]. But it was not until 1963 that Tangherlini made it clear that the 4D Schwarzschild solution could only be embedded in a flat space if the latter had a dimensionality $N = 6$ or higher [5]. In the meantime, Dirac in 1935 had employed an embedding of 4D deSitter space in 5D to give a neat classification of the energies, momenta and masses of particles [6]. Something similar was done by Robertson in 1968 for the standard cosmological solutions of Einstein's equations, including the deSitter universe, which would later be taken as the basis for inflationary cosmology [7]. In 1988, Ponce de Leon found a broad class of 5D metrics which reduced to the standard Friedmann-Robertson-Walker (FRW) cosmologies on hypersurfaces where the fifth coordinate was held fixed [8]. To his surprise, Wesson discovered in 1994 that the Ponce de Leon metrics were *flat* in 5D, first by computer work and then by deriving the algebraic coordinate transformations to 5D Minkowski space $M_5$ [9]. The cosmological side of the embedding saga was finally laid to rest by Lachieze-Rey in 2000 [10], who showed algebraically that *all* of the standard 4D FRW universe models are isometric to $M_5$.

The situation for solutions of Einstein's equations with less symmetry than the cosmological FRW ones remained unclear, however, for many years. Workers were aware that familiar 4D solutions like the Schwarzschild-deSitter one could not be embedded in a *flat* 5D manifold; but there was continued interest in finding some kind of embedding that was algebraically tractable and physically rewarding. It was with this background that Mashhoon in 1994 proposed 5D canonical space [11]. The appellation was suggested by the simple algebraic form of the metric, and by the fact that it led to drastic simplification in the 5D field equations and their corresponding equations of motion (see below). In crude terms, the simplification afforded by the 5D canonical metric is akin to how a 3D metric in Cartesian coordinates *xyz* becomes more transparent for certain situations when expressed in polar coordinates *rθϕ* (this is why problems in geophysics and astrophysics are usually treated in terms of the latter coordinates rather than the former). The general form of 5D canonical space consists of the square of the extra coordinate (divided by a length to preserve physical dimensions) multiplied onto the usual 4D interval, plus an extra flat piece. This 5D form is algebraically general, provided the 4D interval (metric tensor) is allowed to depend on the extra coordinate. It is arrived at by using the 5 coordinate degrees of freedom to eliminate the off-diagonal (electromagnetic) potentials, and to set the magnitude of the extra diagonal potential to unity. This last condition is rather restrictive, and similar to the one used by Kaluza and Klein in the early years of 5D physics. However, it is acceptable, provided it is realized that thereby all of the extra physics associated with the fifth dimension is concentrated into the extra coordinate and its presence as a quadratic factor multiplied onto spacetime. It is therefore necessary to keep an open mind about the meaning of the extra coordinate when working out problems in 5D using the canonical metric (see Section 3). However, the introduction of this form led to the rapid solution of a number of long-standing issues in higher-dimensional physics [12]. This is especially true for the case where the 4D part of the space (metric tensor) does *not* depend on the extra coordinate. The 5D metric in this case is sometimes called "pure canonical". For this case, the theorem quoted at the beginning of this paper holds: *any* solution of Einstein's 4D equations without ordinary matter can be expressed as a 5D metric with the pure-canonical form. This enables a better understanding of the 4D solutions of general relativity, by exploring their 'surroundings' in the extra dimension, which if it is accepted as 'real' also leads to extra physics.

Following the proposition of canonical space in 1994, there was renewed interest in 5D embeddings. Alternatives were investigated, but much of this work is of little interest, because nowadays it is possible to prove the uniqueness of the pure-canonical form $C_5^*$ in short order (a succinct proof will be given in the first segment of Section 3 below). However, there was an interesting byproduct of these investigations. In 1995, the first of a series of papers was published by Tavakol and coworkers [13], which recalled an old theorem from 1926 by Campbell [14]. In that year, Campbell sketched the conditions under which an $N$D Riemannian manifold could be embedded in a similar $(N + 1)$D manifold. The full proof of what became known as Campbell's theorem was left to Magaard in his Ph.D. thesis of 1963 [15]. A more direct proof, using modern methods, was given in 2003 by Seahra and Wesson [16]. The importance of the Campbell-Magaard theorem should not be underestimated. Traditionally, general relativity has been viewed as a stand-alone theory of 4D physics, which while mainly concerned with gravitation might one day be amalgamated with quantum theory to provide a unified theory. A unified theory, that is, in *four* dimensions. The resurrection of the Campbell-Magaard theorem reminded workers of something which should have been obvious throughout academic history: a theory in $N$D is necessarily linked to a wider theory in $(N+1)$D, unless there is some fiat making $N$ (= 4) somehow special. In the absence of such a fiat, and recognizing that Einstein's equations can actually be formulated in *any* number of dimensions, attention turned to the question of the 5D field equations.

The field equations in any $N$D version of general relativity are necessarily based on the Ricci tensor. Contracting this gives the Ricci scalar, which at any point in the $N$D manifold is a scalar measure of curvature, or in physical terms a measure of energy density. From the Ricci tensor and its scalar, it is possible (if desired) to form the Einstein tensor, which by algebraic construction has zero divergence, suggesting its relation to conservation laws. These things are elementary, but should be considered in the passage from (say) 4D to 5D. The 4D field equations are commonly written in the form

$$R_{\alpha\beta} - (R/2) g_{\alpha\beta} + \Lambda g_{\alpha\beta} = \kappa T_{\alpha\beta} \qquad (\alpha, \beta = 0,123) \quad . \tag{1}$$

Here the energy-momentum tensor $T_{\alpha\beta}$ is coupled to the geometry via a constant $(\kappa)$, which by a choice of units may be rendered unity. However, since the components of the metric tensor $g_{\alpha\beta}$ are potentials, the term $\Lambda g_{\alpha\beta}$ involving the cosmological constant essentially sets a zero point for energy. This is unlike other areas of physics, where only the *difference* of energies is considered observable. Sometimes, this problem is circumvented by including the $\Lambda g_{\alpha\beta}$ term in $T_{\alpha\beta}$, as a scalar fluid with the equation of state (pressure plus density) = 0. However, this is specious, because there is an exact cancellation of the numerator in the coupling term with the denominator of the pressure and density of the 'vacuum' defined by (1). That is, the gravitational constant cancels out. These and other problems with the 4D cosmological constant can be avoided by extending the theory from 4D to 5D. In general, the 5D metric will contain not only the 4D metric tensor, but also a scale (describing $\Lambda$) and a scalar field (describing a spin-0 field). In the absence of ordinary matter, Einstein's equations (1) give the 4D Ricci scalar in terms of the cosmological constant as $^4R = 4\Lambda$. This can be used to evaluate the effective (4D) value of $\Lambda$ if $^4R$ can be obtained from the 5D embedding (see Section 3). The 4D equations (1) then read $R_{\alpha\beta} = \Lambda g_{\alpha\beta}$, or if $\Lambda$ is negligible $R_{\alpha\beta} = 0$. These are the equations which are verified from the classical tests of general relativity. Their 5D counterparts are

$$R_{AB} = 0 \quad (A, B = 0, 123, 4) \quad . \tag{2}$$

In these, each component has extra terms compared to 4D, due to the dependency on the extra coordinate $x^4$. These extra terms, involving the extra metric coefficient and the derivatives of the 4D metric coefficients with respect to $x^4$, are important both physically and algebraically. Physically, they allow of the construction of $T_{\alpha\beta}$ in (1), meaning that the energy and momentum content of 4D spacetime is the result of 5D geometry. This is the way in which 5D relativity was arrived at in 1992, in the form of induced-matter or space-time-matter theory [17]. Algebraically, however, it transpires that the embedding of (1) in (2) is *guaranteed* by the Campbell-Magaard theorem. Einstein's 4D field equations (1) with matter are embedded in the (apparently empty) Ricci-flat equations (2). Therefore, 5D relativity is fairly secure in assuming that the field equations are Ricci-flat.

The application of the canonical metric to the field equations (2) makes the latter quite transparent as regards physics. This subject has been well studied in the literature; so here it is sufficient to note that (2) reduce to (1) plus five extra equations. These comprise a set of four conservation relations and a wave equation for the scalar field. In the canonical frame, these extra equations are easily satisfied. The net result, in loose language, is that 5D relativity is similar to standard 4D relativity, except for a quadratic factor in the extra coordinate which is attached to spacetime. It is this factor which determines the embedding.

The history of embeddings, as outlined above, may be summarized as follows. Most solutions of general relativity cannot be embedded in 5D Minkowski space $M_5$. The FRW cosmological models are exceptions, and *are* 5D flat. By contrast, *all* solutions of general relativity can be embedded in the general canonical metric $C_5$. This is by construction; but this embedding also establishes a straight line of reasoning using Campbell's theorem between the 4D Einstein equations (1) and the 5D Ricci equations (2). When the 4D part of the $C_5$ metric does not depend on the extra coordinate (except via the quadratic prefactor), the space is the so-called pure-canonical one $C_5^*$. It is algebraically special, is in general *not* 5D flat, and locally embeds all solutions of general relativity where there is no ordinary matter (i.e. vacuum) but where this is a cosmological constant. This embedding is particularly relevant to the Schwarzschild-deSitter solution [11, 18]. It is a good example of the appropriate use of embeddings in physics. While it may be obvious in retrospect, it should be noted that an arbitrary embedding will not in general 'work', in that it will either conflict with the 5D field equations and/or lead to unphysical components of the 4D energy-momentum tensor. This implies even more respect for those embeddings which *do* work, including the pure-canonical form $C_5^*$ which as will be seen throws new light on the old deSitter solution.

3. <u>Physical Applications of the Canonical Embedding</u>

In this section, by paragraph, the following subjects will be briefly discussed: the uniqueness of the 5D canonical metric; the meaning of the cosmological constant; the 5D null-path and the 4D Klein-Gordon equation; the behaviour of the extra coordinate, and how it leads to variability of the cosmological 'constant'; dynamics when this last parameter dominates; and particle dynamics from a 5D, scale-free perspective.

The notation is standard. The coordinates are $x^0 = t$, $x^{123} = r\theta\phi$ with $d\Omega^2 \equiv \left(d\theta^2 + \sin^2\theta d\phi^2\right)$, and $x^4 = l$ to avoid confusion. The fundamental constants (*c, G, h*) are usually absorbed for ease, except where they are needed for clarity or to obtain a numerical value.

The uniqueness of the canonical metric can be proven most quickly using the field equations $R_{AB} = 0$. The component $R_{44}$ concerns the scalar field, and its precise form may be found in the literature, where it is written out for certain problems in physics. (For example, where the metric coefficients depend on $x^0 = t$, $x^1 = r$ and $x^4 = l$, where the 3D subspace has spherical symmetry for application to astrophysics.) For the present purpose, the metric may be taken in the generic form

$$dS^2 = f(l)ds^2 \pm g(l)dl^2 \quad , \tag{3}$$

where $ds^2 = g_{\alpha\beta}(x^\gamma)dx^\alpha dx^\beta$ defines spacetime. The question is then: What constraint arises from $R_{44} = 0$ between the embedding function $f(l)$ and the scalar field $g(l)$? The answer, after some algebra, is

$$g = \left(\frac{\partial f}{\partial l}\right)^2 \frac{L^2}{f} \quad . \tag{4}$$

Here $L$ is the arbitrary constant of integration which enters from the first integral of the second-order field equation, and is here taken to have the physical dimensions of a length. The constraint (4) makes it straightforward to absorb the scalar field in (3) by a coordinate transformation, thereby fixing the form of the embedding function. The new $x^4$ coordinate is defined up to a second arbitrary constant of integration $(l_0)$, such that $f \sim (l - l_0)^2$ now. Setting the second constant to zero and redefining the first one gives, finally:

$$dS^2 = (l/L)^2 g_{\alpha\beta}(x^\gamma) dx^\alpha dx^\beta \pm dl^2 \quad . \tag{5}$$

This is the (pure) canonical metric as often used in the literature, and is seen to be unique up to coordinate transformations in $x^4 = l$. The length $L$ in (5) may be physically identified in terms of the cosmological constant $\Lambda$, by reducing the 5D field equations to the 4D ones in the absence of ordinary matter. Then $\Lambda = \pm 3/L^2$, where $\Lambda > 0$ for a spacelike extra coordinate and $\Lambda < 0$ for a timelike extra coordinate.

The meaning of the cosmological constant $\Lambda$ can be appreciated by noting that deSitter space of general relativity can be regarded as a 4D pseudosphere of constant curvature, $-1/L^2$ for $\Lambda > 0$ and $+1/L^2$ for $\Lambda < 0$. Alternatively, the 4D metric

$$ds^2 = \left(1 - \Lambda r^2/3\right)dt^2 - \left(1 - \Lambda r^2/3\right)^{-1}dr^2 - r^2 d\Omega^2 \tag{6}$$

can be mapped to the 5D metric

$$dS^2 = dT^2 - \left(dX^2 + dY^2 + dZ^2\right) \pm dL^2 \quad . \tag{7}$$

Here the signature is $(+----)$ for $\Lambda > 0$ and $(+---+)$ for $\Lambda < 0$. Thus, 4D deSitter space is isometric to 5D Minkowski space. The physical cosmological constant $\Lambda$ measures the radius of curvature $L$ which necessarily appears in the algebraic passage from flat 5D to curved 4D. For $\Lambda < 0$, the mapping of deSitter space to a pseudosphere means that a line around the centre can repeatedly traverse the circumference $2\pi L$. This is sometimes seen as a violation of 4D causality since events repeat with a period $2\pi L/c$. If desired, this behaviour can be avoided by supposing that events occur on different surfaces of a tightly-packed 'scroll'. However, from the 5D perspective, this behaviour is natural, as will be seen below.

The null-path in 5D with $dS^2 = 0$ in the canonical metric (5) corresponds to the timelike path in 4D with $ds^2 > 0$ of a massive particle [19]. That is, in 5D *all* particles are in causal contact in the same way as photons in 4D. The nature of the path in 4D depends on the sign of $\Lambda$. For $\Lambda > 0$, the path wanders away from an $l$-hypersurface according to $l = l_* \exp(\pm s/L)$, which is slow if $\Lambda = 3/L^2$ has its (small) cosmological value. For $\Lambda < 0$, the path oscillates around the 4D hypersurface of spacetime, according to

$$l = l_* e^{\pm is/L} \quad , \quad L = \left(3/|\Lambda|\right)^{1/2} \quad . \tag{8}$$

Here $l_*$ is the amplitude of the wave, whose wavelength is $L$. The sign choice merely reflects the reversibility of the motion. A more detailed investigation of the geodesics associated with the canonical metric (5) can be made, based on the variation $\delta\left[\int dS\right] = 0$ around $S = 0$. The 4D motion is identical to that of general relativity for a particle in vacuum, that is if $g_{\alpha\beta} = g_{\alpha\beta}(x^\gamma \text{ only})$. Otherwise, it may be shown that the form (3) with $g_{\alpha\beta} = g_{\alpha\beta}(x^\gamma, l)$ generates matter fields which modify the motion. The motion in the extra dimension for the (pure) canonical metric is a second-order ordinary differential equation for the path $l = l(s)$. This equation is satisfied by the wavelike solution (8), derived above directly from the metric. However, the full form of the equation for the extra geodesic in 5D may be shown by some algebra to be the same as a familiar equation in 4D [20]. Namely, the Klein-Gordon equation of wave mechanics:

$$\Box^2 \psi + m^2 \psi = 0 \quad . \tag{9}$$

Here $\Box^2 \psi \equiv g^{\alpha\beta} \psi_{,\alpha;\beta}$ where a comma denotes the partial derivative and the semicolon denotes the conventional covariant derivative. The symbol $m$ denotes the rest mass of the test particle whose wave function is $\psi = \psi(x^\gamma) = \psi(s)$. The algebraic equivalence between the 5D path (8) and the 4D wave equation (9) has a physical interpretation: the extra parameter $l$ in the canonical metric (5) for a null 5D path plays the *dual* role of extra coordinate and wave function.

The behaviour of the extra coordinate is easier to understand by referring back to the previous discussion of the 4D deSitter metric and its 5D embedding. The deSitter so-lution of general relativity with $\Lambda < 0$ is topologically equivalent to a pseudosphere with radius $L = \left(3/|\Lambda|\right)^{1/2}$, which is closed and whose surface $s$ effectively defines spacetime. Wave-particle duality is implicit to this model, because a particle orbits around the sphere, oscillating about the spacetime surface according to (8). The precise correspondence between algebra and physics involves the speed of light $c$ and Planck's constant $h$ with its modified form $\hbar \equiv h/2\pi$. The algebraic circumference of the deSitter sphere is $2\pi L$. The physical dimension associated with the particle is the Compton wavelength $h/mc$. Setting these equal gives $2\pi L = h/mc$ or $L = \hbar/mc$. This is for the fundamental mode of the wave; but clearly overtones are allowed with wavelengths less by a factor $1/n$ where $n$ is an integer. In this way, it becomes apparent that the 4D deSitter solution can be interpreted in 5D as a trapped wave running around a sphere with radius $\hbar/mc$ where $m$ is the mass of the particle. A consequence of this model is that the cosmological 'constant' is actually a local parameter, and in fact proportional to the square of the particle mass.

With $\Lambda$ variable from system to system, the cosmological-'constant' problem is resolved, at least in principle. This problem consists basically in the mismatch of the small value of $\Lambda$ as determined by cosmology and the large magnitude as determined by particle physics [21, 22]. The discrepancy may be as big as $10^{120}$. However, if $\Lambda$ is a local parameter, then Einstein's equations should be applied system by system each with the appropriate value of this parameter. The cosmological 'constant' should not be assumed to have a unique, universal value. That $\Lambda$ may have a locally-determined value can be appreciated by considering again the embedding of a 4D system in the 5D canoni-cal metric (5). The 4D Ricci scalar for that metric can be calculated in terms of the embedding function, and is $^4R = \pm 12/l^2$ depending on the signature. In the absence of ordinary matter, it is usual to write $^4R = 4\Lambda$, so $\Lambda = \pm 3/l^2$. This is the value determined extrinsically, from the embedding. On the hypersurface $l = L$ of spacetime, $\Lambda = \pm 3/L^2$, agreeing with the result noted before. This is the value determined intrinsically, from Einstein's equations applied in spacetime. The extrinsic and intrinsic values of $\Lambda$ are complementary.

It is instructive, given the preceding comments, to ask about the 4D dynamics of a system with a very large value of $|\Lambda|$. An approximate calculation is sufficient to bring out the main features, though for concreteness attention is focused on the 3D spherically-symmetric system (6) embedded in the 5D metric (5). Then the acceleration of a test particle in the radial direction is

$$\frac{d^2 r}{ds^2} = \frac{\Lambda r}{3} = \pm \frac{r}{L^2} \quad . \tag{10}$$

Here the upper sign is for $\Lambda > 0$ and leads back to Hubble's law for galaxies. The lower sign is for $\Lambda < 0$ and may be relevant to subnuclear dynamics. The latter case is particularly instructive. A purely radial orbit governed by (10) simply oscillates through the (3D) centre of the system with $r = r_* \sin(s/L)$ where $r_*$ is a constant. (There is no acceleration at the centre or asymptotic freedom.) But a circular orbit can be stable if there is a balance between the centrifugal acceleration due to the azimuthal velocity $v_\phi$ and the attraction due to $\Lambda < 0$, thus:

$$\frac{v_\phi^2}{r} = \frac{|\Lambda|c^2 r}{3} = \frac{c^2 r}{L^2} \quad . \tag{11}$$

Presumably the angular momentum of the test particle is quantized in units of the reduced Planck constant, so if the mass is $m$ then $mrv_\phi = n\hbar$ or $v_\phi = n\hbar/mr$ where $n$ is an integer. Combining this with (11) gives the radius of the $n$th. orbit as

$$r_n = \left(\frac{n\hbar L}{mc}\right)^{1/2} \quad . \tag{12}$$

The magnitude of the energy associated with this orbit is

$$E_n = \int_0^{r_n} m\left(\frac{|\Lambda|c^2 r}{3}\right) dr = \frac{mc^2 r_n^2}{2L^2} = \frac{nc\hbar}{2L} \quad , \tag{13}$$

where $r_n$ has been eliminated using (12). The difference in energy between adjacent orbits, and its numerical size for a typical potential well of order $L \approx 10^{-12}$ cm, are:

$$\Delta E \equiv |E_{n+1} - E_n| = \frac{c\hbar}{2L} \approx 1\,\text{GeV} \quad . \tag{14}$$

This is of the same order as the rest-mass energy of many particles, including the proton. It should also be noted that the energy in (13) or (14) is independent of the rest mass of the test particle, and so obeys the Equivalence Principle. It is apparent that the motion of a test particle under the influence of a large $|\Lambda|$ as per equations (10)-(14) is physically reasonable.

Particles in general are conventionally assumed to obey the Klein-Gordon equation (9) if their motions are relativistic. That equation is commonly derived by applying time and space operators for the energy $E$ and 3-momentum $p$ to the metric, or equiva-lently to the energy/momentum relation $E^2 - p^2 = m^2$. But, as noted before, the 4D Klein-Gordon equation is algebraically equivalent to the extra component of the geodesic in 5D dynamics. It is convenient that the 5D geodesic equation splits into a 4D relation which is identical to the spacetime geodesic equation of general relativity, plus an extra relation which is equivalent to the Klein-Gordon equation of standard wave mechanics. However, this split is due to the human propensity to separate spacetime and the extra dimension in 5D dynamics. If desired, it is possible to replace the 4D Klein-Gordon equation in the 4D wave function $\psi_4$ by a 5D wave equation in a 'super' wave-function $\psi_5$. By analogy with (9), this reads

$$\Box_5^2 \psi_5 = 0 \quad . \tag{15}$$

In the weak-field, no acceleration limit of this, with $\psi_5 = \exp[i(Et - px - ml)]$, the conventional equation (9) and the standard relation $E^2 - p^2 - m^2 = 0$ are recovered. (The form of the 5D wave function will depend in general on the form of the 5D metric.) This procedure is admittedly of mainly academic – as opposed to practical – interest. But it illustrates the possibility of replacing a 4D relation (9) with an explicit mass $m$ and the scale attached to it, by a 5D relation (15) in which the mass is implicit and which has no scale. This is relevant to the formulation of particle physics, and notably supersymmetry. The latter is spoilt in 4D by the finite masses of the particles, which are therefore frequently assumed to be boosted from their zero (supersymmetric) values by the Higgs mechanism [23]. By contrast, in a general 5D theory, the masses are related to the scalar field, and the relations describing this and the other fields (gravitation and electromagnet-ism) are scale-free. Indeed the 5D field equations $R_{AB} = 0$, with which the present account began, are scale-free.

4. Conclusion

Five-dimensional Kaluza-Klein theory was invented in the 1920s but effectively went into hibernation until the 1990s. Then, new versions of 5D relativity were proposed on physical grounds. In 1992, induced-matter or space-time-matter theory was proposed as a means of explaining mass and energy as geometrical effects, an approach that was furthered physically by the introduction of the canonical metric in 1994, and given an algebraic basis by the rediscovery of Campbell's theorem in 1995. Also, in 1998 membrane theory was proposed by two different groups, motivated by the wish to understand particle masses and using an exponential in the extra coordinate (rather than a quadratic), which came to be known as the warp metric. This alternative theory has not been dealt with here, partly because it has its own literature and partly because good reviews are already available [24]. Both approaches to 5D relativity are based on the appropriate use of embeddings, and because they pass smoothly to general relativity in the 4D limit they both agree with observations. This correspondence is preserved by the assumption, introduced in 2001, that *all* particles move on null paths in 5D. History, as reviewed in Section 2, therefore leaves the theory as a coherent whole, in which 4D general relativity is locally and smoothly embedded in a larger, causally-connected 5D world.

Newer results, as reviewed in Section 3, have focused on deSitter space. That old 4D solution is now seen as an example of 5D conformal space. With a negative cosmological constant, deSitter space describes a spherical, closed surface with a radius $L = (3/|\Lambda|)^{1/2}$. Identifying this surface with spacetime, a test particle of mass *m* oscillates about the surface in accordance with the 4D Klein-Gordon equation. However, in the 5D picture, the cosmological 'constant' is not universal, but a local parameter whose magnitude is proportional to the square of the particle mass. This prescription is a major consequence of the application of the canonical embedding as it is presently understood.

The 5D perspective offers, of course, the promise of many more insights to 4D physics.

Acknowledgement

Thanks go to various members of the S.T.M. consortium, whose papers are compiled on the website http://astro.uwaterloo.ca/~wesson. This work was partially supported by N.S.E.R.C.